\documentstyle [12pt,eqsecnum,aps,amsfonts] {revtex}
\input epsf
\newcommand{\beq}{\begin{equation}}
\newcommand{\eeq}{\end{equation}}
\newcommand{\beqs}{\begin{eqnarray}}
\newcommand{\eeqs}{\end{eqnarray}}

\catcode`@=11
\@addtoreset{equation}{section}
\@addtoreset{equation}{subsection}
\def\theequation{\ifnum\value{section}=0 \arabic{equation}\ignorespaces
\else \ifnum\value{section}=-1 A.\arabic{equation}\ignorespaces
\else \ifnum\value{subsection}=0 \thesection.\arabic{equation}\ignorespaces
\else \thesection.\arabic{subsection}.\arabic{equation}\ignorespaces
                           \fi
                      \fi
                 \fi}
\catcode`@=12

\begin{document}
\draft

\baselineskip 6.0mm

\title{Ground State Degeneracy of Potts Antiferromagnets on 2D Lattices:
Approach Using Infinite Cyclic Strip Graphs}

\author{Robert Shrock$^{(a)}$\thanks{email: robert.shrock@sunysb.edu}
\and Shan-Ho Tsai$^{(b)}$\thanks{email: tsai@hal.physast.uga.edu}}

\address{(a) \ Institute for Theoretical Physics \\
State University of New York \\
Stony Brook, N. Y. 11794-3840}

\address{(b) \ Department of Physics and Astronomy \\
University of Georgia \\
Athens, GA 30602}

\maketitle

\vspace{4mm}

\begin{abstract}

The $q$-state Potts antiferromagnet on a lattice $\Lambda$ exhibits 
nonzero ground state entropy $S_0=k_B \ln W$ for sufficiently large $q$ 
and hence is an exception to the third law of thermodynamics.  An outstanding
challenge has been the calculation of $W(sq,q)$ on the square (sq) lattice.  
We present here an exact calculation of $W$ on an infinite-length 
cyclic strip of the square lattice which embodies the expected analytic 
properties of $W(sq,q)$.  Similar results are given for the kagom\'e lattice.  

\end{abstract}


\vspace{16mm}

\pagestyle{empty}
\newpage

\pagestyle{plain}
\pagenumbering{arabic}
\renewcommand{\thefootnote}{\arabic{footnote}}
\setcounter{footnote}{0}

    Nonzero ground state entropy, $S_0 \ne 0$, is an important subject in
statistical mechanics as an exception to the third law of thermodynamics (e.g.,
\cite{cw}). This is equivalent to a ground state degeneracy per site $W > 1$,
since $S_0 = k_B \ln W$.  The $q$-state Potts antiferromagnet (AF)
\cite{potts,wurev} exhibits nonzero ground state entropy (without frustration)
for sufficiently large $q$ on a given lattice $\Lambda$, or more generally, a
given graph $G$, and serves as a valuable model for the study of this
phenomenon.  The zero-temperature partition function of the above-mentioned
$q$-state Potts AF on $G$ satisfies $Z(G,q,T=0)_{PAF}=P(G,q)$, where $P(G,q)$
is the chromatic polynomial (in $q$) expressing the number of ways of coloring
the vertices of the graph $G$ with $q$ colors such that no two adjacent
vertices have the same color \cite{rtrev,chi}.  Thus, $W(\{G\},q)= \lim_{n \to
\infty} P(G,q)^{1/n}$, where $n=v(G)$ is the number of vertices of $G$
\cite{noncomm,w} and $\{G\}=\lim_{n \to \infty}G$.  $W(\{G\},q)$ has been
calculated exactly for the triangular lattice \cite{baxter} and various
families of graphs \cite{w}, \cite{wc}-\cite{wa2}. The special values for the
square (sq) and kagom\'e (kg) lattices $W(sq,3)$ \cite{lieb} and $W(kg,3)$
(which can be extracted from \cite{baxter70,baxter}) are also known.  However,
aside from the triangular case, the exact calculation of $W(\Lambda,q)$ for
general $q$ on lattices $\Lambda$ of dimensionality $d \ge 2$ remains an
outstanding challenge.  In this work we report exact calculations of $W$ on
infinite-length, finite-width strips of the square and kagom\'e lattices that
exhibit the analytic properties expected for the $W$ functions on the
respective full 2D lattices and, in this sense, constitute the closest exact
results that one has to these $W$ functions.

Let us describe these analytic properties.  Denote $\lim_{n \to \infty} G =
\{G\}$.  Since $P(G,q)$ is a polynomial, one can generalize $q$ from ${\mathbb
Z}_+$ to ${\mathbb R}$ and indeed ${\mathbb C}$. $W(\{G\},q)$ is a real
analytic function for real $q$ down to a minimum value, $q_c(\{G\})$
\cite{w,p3afhc}.  For a given $\{G\}$, we denote the continuous locus of
non-analyticities of $W$ as ${\cal B}$.  This locus ${\cal B}$ forms as the
accumulation set of the zeros of $P(G,q)$ (chromatic zeros of $G$) as $n \to
\infty$ \cite{bds,bkw,read91,w} and satisfies ${\cal B}(q)={\cal B}(q^*)$.  In
cases where ${\cal B}$ serves as a natural boundary, dividing the $q$ plane
into different regions, $W$ has different analytic forms in these different
regions.  The point $q_c$ is the maximal point where ${\cal B}$ intersects the
real axis, which can occur via ${\cal B}$ crossing this axis or via a line
segment of ${\cal B}$ lying along the axis.  The chromatic polynomial $P(G,q)$
has a general decomposition as $P(G,q) = c_0(q) + \sum_j c_j(q)(a_j(q))^{t_j
n}$ where the $a_j(q)$ and $c_{j \ne 0}(q)$ are independent of $n$, while
$c_0(q)$ may contain $n$-dependent terms, such as $(-1)^n$, but does not grow
with $n$ like $(const.)^n$ with $|const.| > 1$, and $t_j$ is a $G$-dependent
constant.  A term $a_\ell(q)$ is ``leading'' $(\ell)$ if it dominates the $n
\to \infty$ limit of $P(G,q)$.  The locus ${\cal B}$ occurs where there is an
abrupt nonanalytic change in $W$ as the leading terms $a_\ell$ changes; thus
the locus ${\cal B}$ is the solution to the equation of degeneracy of
magnitudes of leading terms.  Hence, $W$ is finite and continuous, although
nonanalytic, across ${\cal B}$.

 From exact calculations of $W$ on a number of families of graphs we have
inferred several general results on ${\cal B}$: (i) for a graph $G$ with
well-defined lattice structure, a sufficient condition for ${\cal B}$ to
separate the $q$ plane into different regions is that $G$ contains at least one
global circuit, defined as a route following a lattice direction which has the
topology of the circle $S^1$ and a length $\ell_{g.c.}$ that goes to infinity
as $n \to \infty$ \cite{strip,suff}.  For a $d$-dimensional lattice graph, the
existence of global circuits is equivalent to having periodic boundary
conditions (BC's) in at least one direction. Further, (ii) the general
condition for a family $\{G\}$ to have a locus ${\cal B}$ that is noncompact
(unbounded) in the $q$ plane \cite{wa} shows that a sufficient (not necessary)
condition for $\{G\}$ to have a compact, bounded locus ${\cal B}$ is that it is
a regular lattice \cite{wa,wa3,wa2}.  The third and fourth general features are
that for graphs that (a) contain global circuits, (b) cannot be written in the
form $G=K_p + H$ \cite{wc,graphdef}, and (c) have compact ${\cal B}$, we have
observed that ${\cal B}$ (iii) passes through $q=0$ and (iv) crosses the
positive real axis, thereby always defining a $q_c$.

  From exact calculations of $W$ for a number of infinite-length, finite-width
(homogeneous) strips of 2D lattices with free boundary conditions in the
longitudinal direction (and free or periodic BC's in the transverse direction)
\cite{strip,hs}, it is found that the resultant loci ${\cal B}$ consist of arcs
(and possible real line segments) which, although compact, do not separate the
$q$ plane into different regions, do not pass through $q=0$ and, for the arcs,
do not necessarily intersect the real $q$ axis.  These calculations showed that
as the strip width $L_y$ increases, the complex-conjugate (c.c.) arcs
comprising ${\cal B}$ tend to elongate so that the gaps between them decrease,
and the left endpoints of the c.c. arcs nearest to $q=0$ move toward this
point, thereby leading to the inference that in the limit $L_y \to \infty$,
these arcs will close to form one or more regions, and ${\cal B}$ will pass
through $q=0$ and will cross the positive real axis at one or more points,
thereby defining a $q_c$.  In turn, this motivates the conclusion that the
properties (i)-(iv) hold for $W(\Lambda,q)$ and ${\cal B}$ on a lattice
$\Lambda$ (in the thermodynamic limit, independent of the boundary conditions
used).  The advantage of cyclic strip graphs is that these properties are
present for each finite $L_y$ rather than only being approached in the limit
$L_y \to \infty$ as for open strips.

Our method for obtaining exact $W$ functions that exhibit the analytic 
properties expected for $W$ on a 2D lattice is as follows.
We calculate $P(G_\Lambda,q)$ on $L_x \times L_y$ strips of the lattice
$\Lambda$ with periodic (i.e., cyclic) BC's in the longitudinal ($L_x$)
direction, then take $L_x \to \infty$ and calculate $W$ and the resultant
${\cal B}$. By construction, these $W$ functions and the associated loci ${\cal
B}$ embody the four general properties given above.  For each strip, the
exterior of ${\cal B}$ in the $q$ plane, denoted as the region $R_1$, is the
maximal region into which one can analytically continue $W$ from the real
interval $q > q_c$.  The calculation of $W$ for a cyclic strip of a given width
is considerably more difficult and the result more complicated than that for
the open strip (free $L_x$ BC) of the same width; the value of the cyclic
strips is that the resultant $W$ exhibits the analytic features of the full 2D
$W$ function.  The boundary condition in the transverse direction is not
important for these results since the width is finite; for simplicity we use
free transverse BC's.

We use an extension of the generating function method of Ref. \cite{strip} from
open to cyclic strip graphs $G_\Lambda$.  The generating function
$\Gamma(G_\Lambda,q,x)$ yields the chromatic polynomials for finite-length
strips of $\Lambda$ as the coefficients in its Taylor series expansion in the
auxiliary variable $x$ about $x=0$.  Here, $\Gamma(G_\Lambda,q,x) = {\cal
N}(G_\Lambda,q,x)/{\cal D}(G_\Lambda,q,x)$, where ${\cal N}$ and ${\cal D}$ are
polynomials in $x$ and $q$ (with no common factors).  The degrees of these, as
polynomials in $x$, are denoted $j_{max}=deg_x({\cal N})$ and
$k_{max}=deg_x({\cal D})$.  The ${\cal N}$ are not needed here (they will be
given elsewhere) since $W$ and ${\cal B}$ are determined completely by ${\cal
D}$, independent of ${\cal N}$ \cite{strip}.  For a particular $G_\Lambda$,
writing ${\cal D}=\prod_{j=1}^{j_{max}}(1-\lambda_j x)$, $W$ is given in region
$R_1$ and $|W|$ in other regions \cite{mag} by $W=(\lambda_{max})^t$ and
$|W|=|\lambda_{max}|^t$, where $\lambda_{max}$ denotes the $\lambda$ in $P$
with maximal magnitude in the respective region and $t=L_x/n=1/L_y$ for the
square strip and 1/5 for the kagom\'e strip considered here.

We first consider cyclic strips of the square lattice.  For $L_y=1$, ${\cal B}$
consists of the unit circle $|q-1|=1$ so that $q_c=2$, and $W=q-1$ for $q \in
R_1$.  For $L_y=2$, from the known $P$ \cite{bds}, we found that ${\cal B}$
separates the $q$ plane into four regions, $q_c=2$, and $W=(q^2-3q+3)^{1/2}$
for $q \in R_1$ \cite{w}.  We have calculated the generating function for the
$L_y=3$ case.  This has $j_{max}=8$ and $k_{max}=10$ and is considerably more
complicated than the $L_y=3$ open strip, where $j_{max}=1$ and $k_{max}=2$.
For ${\cal D}$ we find 
\beqs 
& & {\cal D}(sq(L_y=3),q,x) =
(1+b_{sq,11}x+b_{sq,12}x^2) (1+b_{sq,21}x+b_{sq,22}x^2+b_{sq,23}x^3) \times \cr
& & (1+x)[1+(q-2)^2x][1-(q-2)x][1-(q-4)x][1-(q-1)x]
\label{dsq}
\eeqs
where 
\beq
b_{sq,11}=-(q-2)(q^2-3q+5)
\label{bsq11}
\eeq
\beq
b_{sq,12}=(q-1)(q^3-6q^2+13q-11)
\label{bsq12}
\eeq
\beq
b_{sq,21}=2q^2-9q+12
\label{bsq21}
\eeq
\beq
b_{sq,22}=q^4-10q^3+36q^2-56q+31
\label{bsq22}
\eeq
\beq
b_{sq,23}=-(q-1)(q^4-9q^3+29q^2-40q+22)
\label{bsq23}
\eeq 

\begin{figure}
\centering
\leavevmode
\epsfxsize=3.5in
\begin{center}
\leavevmode
\epsffile{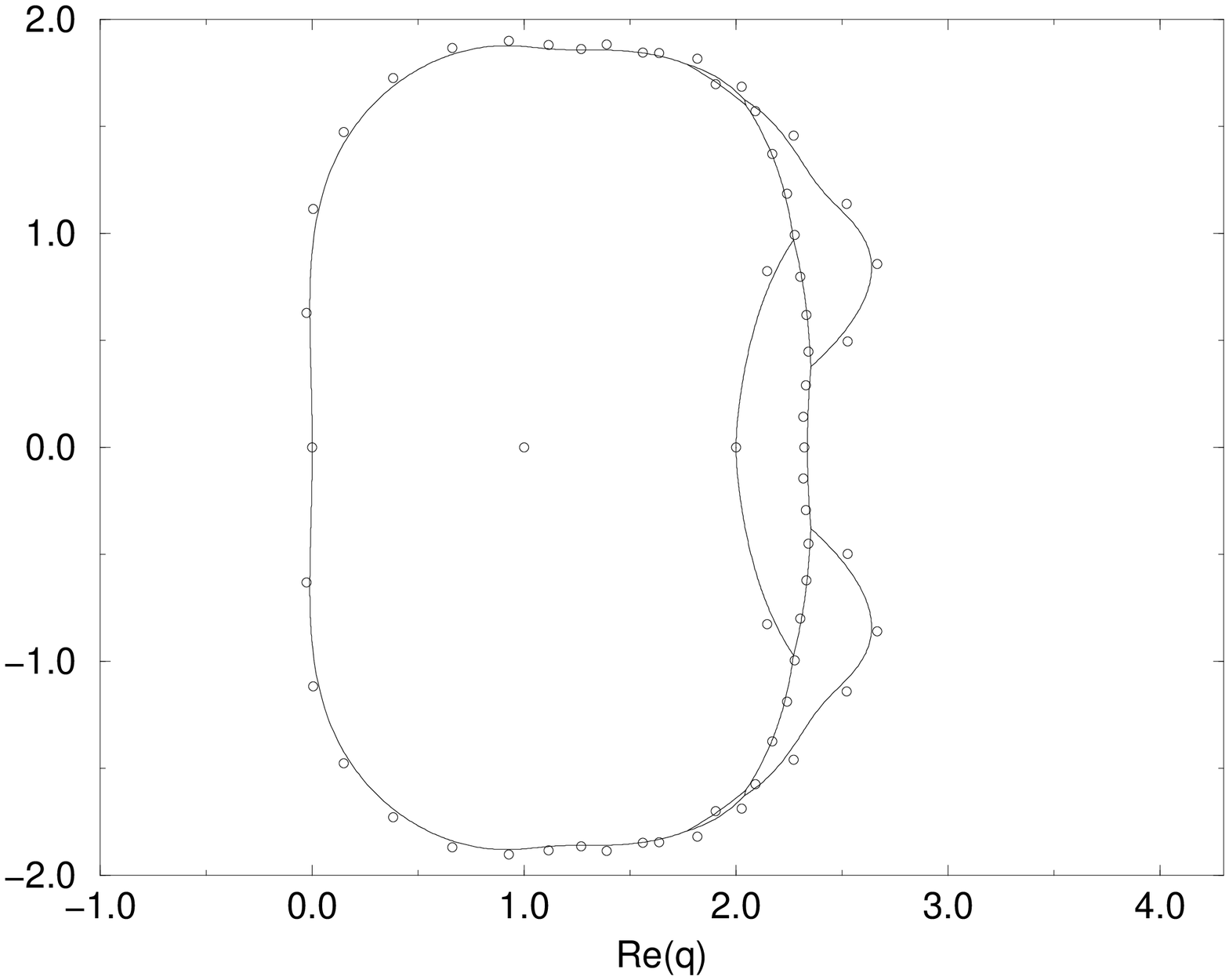}
\end{center}
\vspace{-2cm}
\caption{\footnotesize{Locus ${\cal B}$ for $W$ for the $\infty \times 3$
cyclic strip of square lattice.  Chromatic zeros for $L_x=20$ (i.e., $n=60$) 
are also shown.}}
\label{clad3}
\end{figure}
The boundary ${\cal B}$ is shown in Fig. \ref{clad3}.  It divides the $q$ plane
into several regions and crosses the positive real axis at 
$q_c=2.33654$ and $q=2$.  Thus, this $L_y=3$ cyclic strip is the first one
sufficiently wide as to yield a value of $q_c$ above $q=2$; indeed, the value
of $q_c$ for this strip is only about 20 \% below the value for the full 2D 
lattice, viz., $q_c=3$ \cite{w}.  In region $R_1$ including the real interval 
$q > q_c$, 
\beqs 
W(\{G_{sq(L_y=3)}\},q \in R_1) & = & 2^{-1/3}\biggl [ (q-2)(q^2-3q+5) +
\nonumber \\ & & \Bigl [(q^2-5q+7)(q^4-5q^3+11q^2-12q+8) \Bigr ]^{1/2} \biggr
]^{1/3}
\label{wsqw2}
\eeqs 
At $q_c$, $W=1.18487$.  In the region that includes the real
interval $2 < q < q_c$, $|W|=|q-4|^{1/3}$.  In the region that
includes the real interval $0 < q < 2$ and in the regions centered at roughly 
$q = 2.4 \pm 0.9i$, $|W|$ is given by the respective maximal cube roots of the
equation $\xi^3+b_{sq,21}\xi^2+b_{sq,22}\xi+b_{sq,23}=0$.
As an algebraic curve, ${\cal B}$ has several multiple points (defined 
as points where several branches of this curve cross intersect). 

We next consider a cyclic strip of the kagom\'e lattice comprised of $m$ 
hexagons with each pair sharing two triangles as adjacent polygons (as in 
Fig. 1 in \cite{strip} for the open strip).  $\Gamma$ has $j_{max}=8$ and 
$k_{max}=9$ as compared with $j_{max}=1$, $k_{max}=2$ for the open strip of the
same width \cite{strip}.  We calculate 
\beqs
{\cal D}(kg(L_y=2),q,x) &=& (1+b_{kg,11}x+b_{kg,12}x^2)(1+b_{kg,21}x+
b_{kg,22}x^2)(1+b_{kg,31}x+b_{kg,32}x^2)\times \cr \cr
& &[1-(q-2)x][1-(q-4)x][1-(q-1)(q-2)^2x]
\label{dkag}
\eeqs
where 
\beq
b_{kg,11}=-(q-2)(q^4-6q^3+14q^2-16q+10)
\eeq
\beq
b_{kg,12}=(q-1)^3(q-2)^3
\eeq
\beq
b_{kg,21}=-q^3+7q^2-19q+20
\eeq
\beq
b_{kg,22}=(q-1)(q-2)^3
\eeq
\beq
b_{kg,31}=11-9q+2q^2
\eeq
\beq
b_{kg,32}=-(q-1)(q-2)^2
\eeq
Define 
$\lambda_{kg,j,\pm}=(1/2)[-b_{kg,j1} \pm (b_{kg,j1}^2-4b_{kg,j2})^{1/2}]$. 
Again, ${\cal B}$ divides the $q$ plane into several regions (Fig. 2).  In 
region $R_1$, $W$ is determined by $\lambda_{kg,1,+}$: 
\beqs
W(\{G_{kg(L_y=2)}\},q) & = & 2^{-1/5}(q-2)^{1/5}\biggl [ q^4-6q^3+14q^2-16q+10
\nonumber \\
& & + \Bigl [q^8-12q^7+64q^6-200q^5+404q^4-548q^3+500q^2-292q+92
  \Bigr ]^{1/2} \biggr ]^{1/5}
\label{wkagw2}
\eeqs 
As is evident from Fig. 2, the value of $q_c$ is within about 10 \% of
the inferred exact value $q_c=3$ for the 2D kagom\'e lattice \cite{w2d}.  It is
impressive that an infinite strip of width $L_y=2$ yields a $q_c$ this close to
the value for the full 2D lattice.

\begin{figure}
\centering
\leavevmode
\epsfxsize=3.5in
\begin{center}
\leavevmode
\epsffile{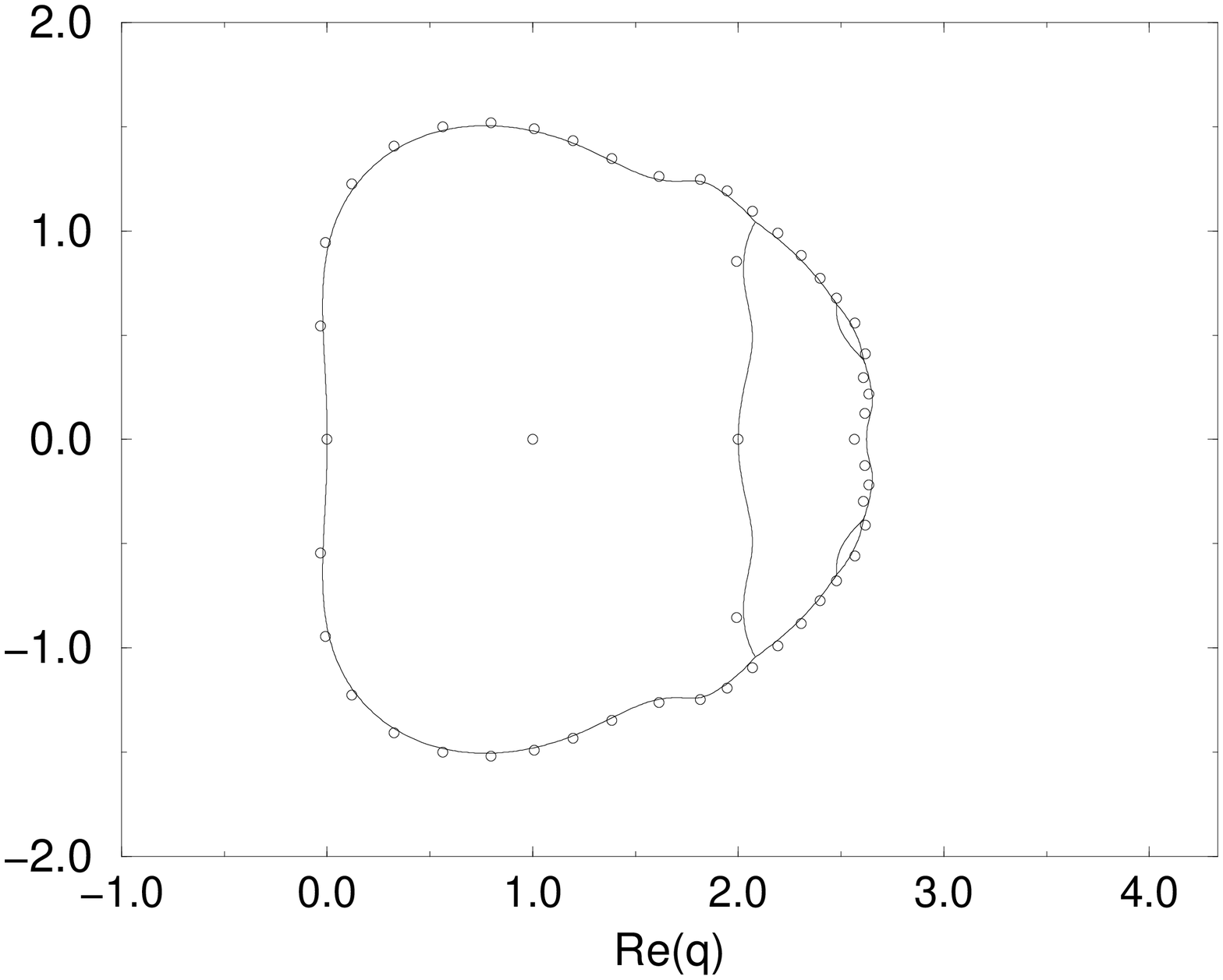}
\end{center}
\vspace{-2cm}
\caption{\footnotesize{Locus ${\cal B}$ for $W$ for $\infty \times 2$ cyclic
strip of the kagom\'e lattice. Chromatic zeros for $m=10$ ($n=50$) are also
shown.}}
\label{cyckag}
\end{figure}

Another interesting
feature of these results is the fact that the chromatic zeros and their
accumulation set ${\cal B}$ contain support for $Re(q) < 0$. This is in
contrast with the situation for strips with free longitudinal BC's \cite{strip}
and provides further support for our earlier conjecture that a necessary
condition for this $Re(q) < 0$ feature is that the graph family have global
circuits.

We have also computed $W$ and ${\cal B}$ for the cyclic strip of
the triangular strip with $L_y=2$.  We find 
${\cal D}=(1-x)[1-(q-2)^2x][1+(2q-5)x+(q-2)^2x^2]$.  ${\cal B}$ separates the
$q$ plane into three regions and crosses the positive real axis at $q_c=3$ and
at $q=2$.  The $q_c$ value for this strip is one unit less than the value 
$q_c=4$ for the full 2D lattice. 

Similar calculations can be carried out for infinite-length
cyclic strips $G_\Lambda$ of greater widths.  Our method can also be
applied to lattices with $d \ge 3$.  To do this, one would use the generating
function method to calculate $P$ for tubes with longitudinal PBC's and
successively larger $(d-1)$-dimensional cross sections.  We believe that this
application, as well as that to other 2D lattices, is promising.

This research was supported in part by the NSF grant PHY-97-22101.

\vfill
\eject


\begin{thebibliography}{99}
\bibitem{chi}{The minimum number of colors needed for this coloring of
$G$ is called its chromatic number, $\chi(G)$.}
\bibitem{noncomm}{At certain special points $q_s$ (typically $q_s=0,1,..,
\chi(G)$), one has the noncommutativity of limits $\lim_{q \to q_s} \lim_{n \to
\infty} P(G,q)^{1/n} \ne \lim_{n \to \infty} \lim_{q \to q_s}P(G,q)^{1/n}$, and
hence it is necessary to specify the order of the limits in the definition of
$W(\{G\},q_s)$ \cite{w}.  As in Ref. \cite{w}, we shall use the first order of
limits here; this has the advantage of removing certain isolated
discontinuities in $W$.}
\bibitem{suff}{Some families of graphs that do not have regular lattice
directions have noncompact loci ${\cal B}$ that separate the $q$ plane into 
different regions \cite{wa,wa3,wa2}.}
\bibitem{graphdef}{The complete graph on $p$ vertices, denoted $K_p$, is the 
graph in which every vertex is adjacent to every other vertex.  The ``join'' of
graphs $G_1$ and $G_2$, denoted $G_1 + G_2$, is defined by adding bonds 
linking each vertex of $G_1$ to each vertex in $G_2$.  Graph families with 
${\cal B}$ not including $q=0$ are given in \cite{wc,wa,wa3,wa2}.}
\bibitem{mag}{For real $q < q_c(\{G\})$, as well as other regions of the $q$
plane that cannot be reached by analytic continuation from the real interval $q
> q_c(\{G\})$, one can only determine the magnitude $|W(\{G\},q)|$
unambiguously \cite{w}.}
\end{thebibliography}
\end{document}